\documentclass[twoside]{article}
\usepackage{fleqn,espcrc2}
\usepackage{graphicx}
\usepackage[figuresright]{rotating}

\newcommand{\AmS}{{\protect\the\textfont2
  A\kern-.1667em\lower.5ex\hbox{M}\kern-.125emS}}

\input{psfig.sty}
\title{(Mis)Understanding the Atmospheric Neutrino Anomaly}
\author{J.M.~LoSecco
\address{University of Notre Dame, Notre Dame, Indiana 46556}}
\begin{document}
\maketitle

\begin{abstract}
The apparent attenuation of muon neutrinos relative to electron neutrinos is
a bit too low to be compatible with the most popular values of $\Delta m^{2}$.
Fits to $R(E_{\nu})$ favor values of $\Delta m^{2}>0.1$ eV$^{2}$.
The fit minimized by the Super Kamioka group in estimating neutrino oscillation
parameters neglects systematic errors.  The fit is dominated by systematic
effects.  The data being combined in recent fits may not be compatible since
there appear to be significant variations in the properties of the data with
time.
A simple two component neutrino oscillation with $\Delta m^{2}$ in the range
of 10$^{-3}$ to 10$^{-2}$ eV$^{2}$ seems unable to account for the
observations.
\vspace{1pc}
\end{abstract}

\section{Introduction}
Atmospheric neutrinos originate from the decay of unstable particles produced
by cosmic ray interactions in the upper atmosphere.  The useful spectrum
runs from about 200 MeV to about 1330 MeV.  But flux extends to
much higher energies.  The Super Kamioka group has studied these out to about
10 GeV.  The higher energy sample is only about 21\% of the size of the
sample below 1330 MeV.

\begin{figure*}
{\mbox{\psfig{file=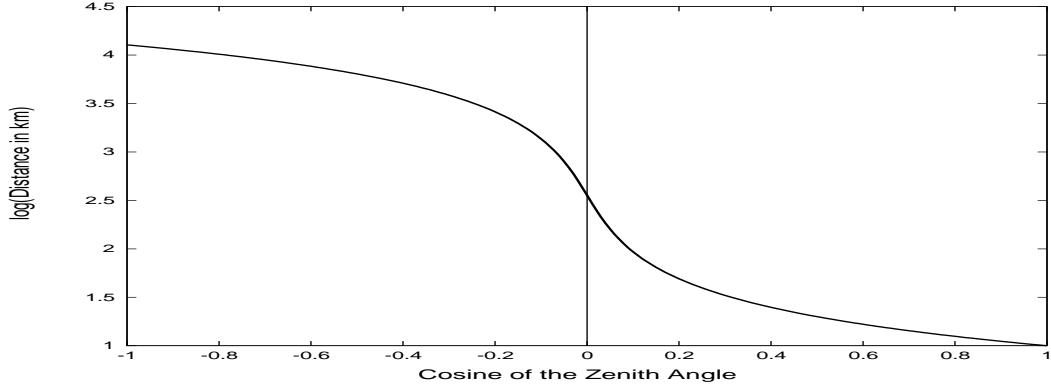,height=2in,width=5.5in}}}
\caption{\label{dist} Log of the flight distance in km as a function of zenith
angle.  Downward going is on the right.}
\end{figure*}

Atmospheric neutrinos approximate the ideal of a two distance neutrino
oscillation experiment.  Figure \ref{dist} shows the log of the flight
length for neutrinos as a function of the cosine of the zenith angle.
Events from below travel on the order of 8000 km.  Those coming from above
travel on the order of 10 km.  There is rather little solid angle at
intermediate distances.  Atmospheric neutrinos provide a good mixture of
electron and muon neutrinos since a significant number of the muons at
these energies
decay before reaching the ground.  Simple arguments suggest that the ratio
of muon to electron neutrinos should be about 2.

To summarize, atmospheric neutrinos are a mixture of muon and electron
neutrinos coming from hadronic and muon decays.
The source is approximately at two distances that differ by
three orders of magnitude.  Most of the flux spans about a factor of 7 in
energy.

The atmospheric neutrino anomaly\cite{haines} is the observation of an
apparent deficiency of muon type atmospheric neutrinos.
The anomaly is believed to be
evidence for neutrino oscillations\cite{kayser}.  But some of the details
and assumptions used in the analysis cast some doubt on the oscillation
hypothesis.  The neutrino sample is being drawn from natural sources which adds
a degree of uncertainty.

Almost all analyses have assumed that the anomaly is restricted to the
muon neutrinos and that the electron neutrino flux is a good normalizer.

\section{R(E$_{\nu}$)}
The ratio of observed to expected neutrino interactions is a measure of the
possible reduction in the neutrino flux that is indicative of neutrino
oscillations.  Due to a significant uncertainty in the overall flux
normalization, as much as 25\%, the quality of the attenuation statistic is
improved by normalizing to something which is better known.  The value of
$R = \frac{((\mu/e)_{Data}}{(\mu/e)_{MC}}$\cite{hirata} which attempts to
normalize the muon neutrino flux to the electron neutrino flux is thought
to have significantly smaller systematic errors, on the order of 5\%.

An advantage that $R$ has over other observables is that it maximizes the
use of the data.  It is insensitive to details of the angular correlation
between neutrino direction and reconstructed lepton direction from the
interaction.  A disadvantage is that one is averaging over two different
baselines so that if only the long distance sample has oscillated significantly
the result will be diluted by the unoscillated sample from the shorter
baseline.

\begin{figure*}
{\mbox{\psfig{file=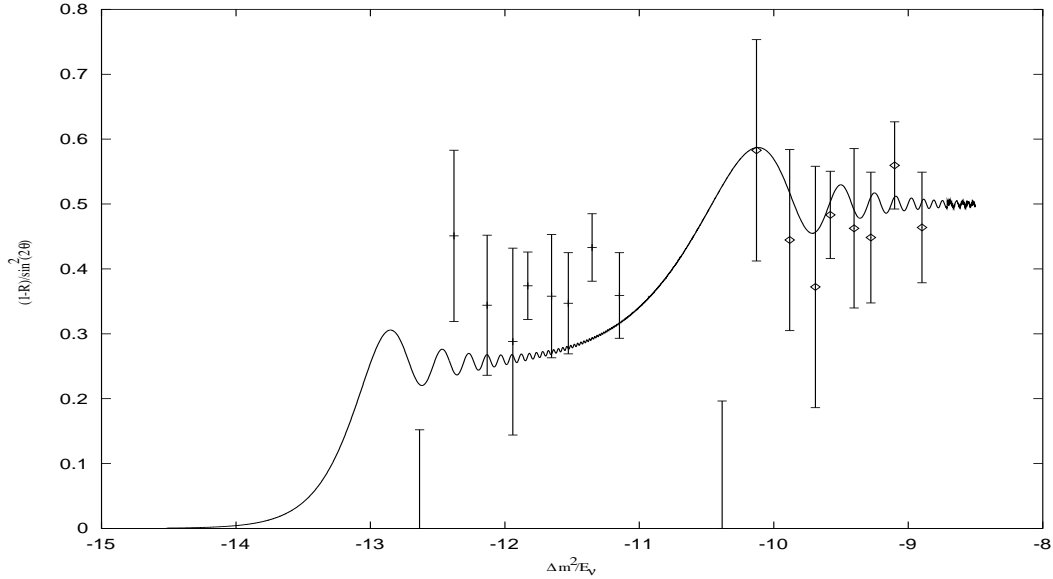,height=3in,width=5.5in,angle=270}}}
\caption{\label{rnu}$\frac{1-R}{\sin^{2}(2 \theta)}$ as a function of the
log$_{10}$ of $\Delta m^{2}/E_{\nu}$ with $\Delta m^{2}$ in eV$^{2}$
and $E_{\nu}$ measured in eV.  (Add 9 to get $E_{\nu}$ measured in GeV).
The points plotted near the center of the figure are from references
[5] and [6] using the $\Delta m^{2}$ from reference
[7].  The points plotted to the right are our fit to $\Delta m^{2}$
using the data of [5] and [6].}
\end{figure*}

\subsection{The Fit to $R(E_{\nu})$}
Figure \ref{rnu} illustrates the Super Kamioka measurement of $R$.
The vertical scale is $\frac{1-R}{\sin^{2}(2 \theta)}$ which combines the
observed data, $R$ with a fit parameter $\sin^{2}(2 \theta)$.
The horizontal axis of figure \ref{rnu} is the log$_{10}$ of
$\Delta m^{2}/E_{\nu}$ with $\Delta m^{2}$ in eV$^{2}$ and $E_{\nu}$ measured
in eV.
The points at
the center of the plot are taken from references \cite{sub} and \cite{mult}
and are plotted with $\sin^{2}(2 \theta)=1$ which is obtained from
the Super Kamioka fit described in the next section.  Since, for these
point $\sin^{2}(2 \theta)=1$ the value of $1-R$ can be read off the graph.
Reference \cite{sub} quotes $R=0.61\pm0.03\pm0.05$ for energies below
1330 MeV.  Reference \cite{mult} quotes $R=0.66\pm0.06\pm0.08$ for the
higher energy sample.

The diamond points in figure \ref{rnu} are the same data but the values
of $\sin^{2}(2 \theta)$ and $\Delta m^{2}$ are a fit to these data points.
The fit yields $\sin^{2}(2 \theta)=0.8$ and $\Delta m^{2}=0.24$ eV$^{2}$.
The points near the center, the Super Kamioka fit, would agree with the
data much better if the points could be lowered.  But to do this would
require $\sin^{2}(2 \theta)=1.5$ which is unphysical.  It is difficult
to assign a confidence level to our fit.  The points share many common
systematic errors.  One would need to quantify these correlated errors
in order to achieve a legitimate confidence interval.  Since we have
plotted $\Delta m^{2}/E_{\nu}$ the higher energy data points are to the
left.  The $\Delta m^{2}$ mass scale is driven by the low value of $R$
measured for the higher energy data points.  A reasonable fit can be obtained
for $\Delta m^{2}>0.1$ eV$^{2}$.  More details of the fit can be found in
\cite{rofnu}.

The curve plotted in figure \ref{rnu} is an integral over the distances
of figure \ref{dist} for $1-R$.  For single component neutrino oscillations
the result is a function of $\Delta m^{2}/E_{\nu}$ and proportional
to $\sin^{2}(2 \theta)$.  The curve illustrates the structure expected from
the two distance scales involved.  At very small $\Delta m^{2}$ all path
lengths are small compared to the oscillation length and so the value of
$R$ is 1.  At very large $\Delta m^{2}$  the oscillation length is small
compared to all path lengths in the problem so that $1-R$ will be
$0.5 \times \sin^{2}(2 \theta)$,  For a range of intermediate values of
$\Delta m^{2}$ one path length is large compared to the oscillation length
and the other is small so that $1-R \approx 0.75 \times \sin^{2}(2 \theta)$,

\begin{figure*}
{\mbox{\psfig{file=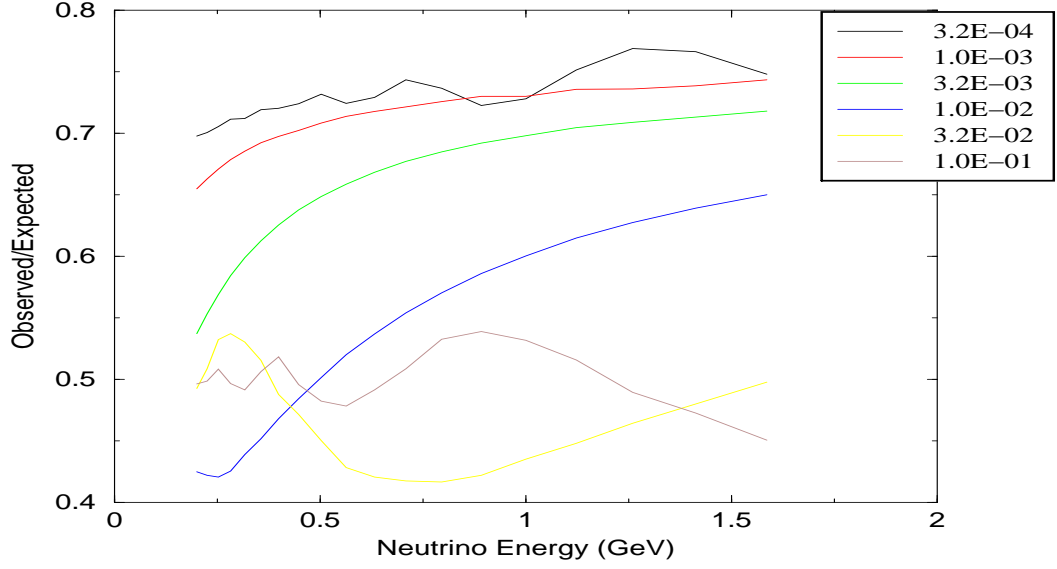,height=3in,width=5.5in}}}
\caption{\label{rofe} A plot of $R(E_{\nu})$ for $\sin^{2}(2 \theta)=1$
and various $\Delta m^{2}$.  $\Delta m^{2}$ runs from
$3.2 \times 10^{-4}$ eV$^{2}$ to $0.1$  eV$^{2}$ as marked in the legend.
These correspond in general to the ordering of the curves from top to bottom.}
\end{figure*}

\subsection{Geomagnetic Effects}
Figure \ref{rofe} illustrates the energy dependence of $R$ for various values
of $\Delta m^{2}$.  The figure was obtained by numerically integrating the
Honda 97 muon neutrino flux\cite{thnks}.  For modest values of $\Delta m^{2}$
the value of $R$ is energy independent over the range indicated.  But it can
never drop much below 0.75.  As $\Delta m^{2}$ rises portions of $R$ drop
to below 0.75 and start to approach 0.5.  Only at values of $\Delta m^{2}$
of about 0.1 eV$^{2}$ is energy independence restored with $R<0.75$.
Figure \ref{rofe} was generated for $\sin^{2}(2 \theta)=1$, its maximum
value.  For smaller values of $\sin^{2}(2 \theta)$ the allowed values of $R$
are greater.  This plot supports the fit of figure \ref{rnu} in that
the only curve that gives an energy independent value for $R(E_{\nu})$
at a value of $R$ below $\approx$0.75 is $\Delta m^{2}=0.1$ eV$^{2}$

The curve of figure \ref{rnu} was done assuming an isotropic flux.
A check using the Honda 97 flux reproduces the results to about 1\%, except
at the lowest energy point, E=250 MeV, where the difference between isotropy
and the calculated flux is less than about 3\% for $\Delta m^{2} \leq 1 \times
10^{-4}$ eV$^{2}$.

\subsection{Physical Interpretation}
The physical significance of a low value of $R$, if interpreted in terms
of only a muon deficiency due to muon neutrino oscillations integrated over
energy and distance, is that both distance scales must participate.
Since the short distance scale is about 3 orders of magnitude shorter than
the large one the favored value for $\Delta m^{2}$ must be comparably
larger.

\section{The Super Kamioka Fit}
\begin{table*}
\begin{tabular}{l|lrrrrrr}
$\alpha$ & overall normalization & & & & & & \\
$\delta$ & E$_{\nu}$ spectral index & & & & & & \\
$\beta_{s}$ & sub-GeV $\mu$/e ratio & & & & & & \\
$\beta_{m}$ & multi-GeV $\mu$/e ratio & & & & & & \\
$\rho$ & relative norm of PC to FC & & & & & & \\
$\lambda$ & L/E$_{\nu}$ & & & & & & \\
$\eta_{s}$ & sub-GeV up-down & & & & & & \\
$\eta_{m}$ & multi-GeV up-down & & & & & & \\
\end{tabular}
\caption{\label{syst}Super Kamioka fit parameters}
\end{table*}

\subsection{Systematic Errors}
The Super Kamioka group has fit data to the neutrino oscillation hypothesis
using:
\[
\chi^{2}(\sin^{2}(2\theta),\Delta m^{2},\vec{\epsilon}) = \sum_{i=1}^{70}
\frac{(N^{i}_{data}-N^{i}_{MC})^{2}}{\sigma^{2}_{i}}
+\sum_{j} \frac{\epsilon_{j}^{2}}{\sigma^{2}_{j}}
\]

$\sigma_{i}$ includes the statistical error for the observed number of events
in the $i$'th bin.  $\sigma_{j}$ is the error on the $j$'th fitted
``systematic'' parameter.  $\sigma_{j}$ is assumed to be known.
$\vec{\epsilon}$ is a vector of 8 systematic
parameters.  The sum is over 70 bins; 5 bins in solid angle, 7 bins in energy
for both electron and muon like events.

The number of Monte Carlo events is estimated from

\[
N^{i}_{MC} = \frac{{\cal L}_{data}}{{\cal L}_{MC}} \times \sum_{MC events} w
\]

Where the $\cal L$ is the relative livetime of the data or Monte Carlo and
\begin{eqnarray*}
w=(1+\alpha) (E_{\nu}^{i}/E_{0})^{\delta} (1+\frac{\eta_{s,m}}{2}
\cos(\Theta))\\
\times f_{e,\mu}(\sin^{2}(2\theta),\Delta m^{2},(1+\lambda)L/E_{\nu})\\
\times
\left\{
\begin{array}{l}
(1-\beta_{s}/2)\\
(1+\beta_{s}/2)\\
(1-\beta_{m}/2)\\
(1+\beta_{m}/2)(1-\frac{\rho}{2}\frac{N_{PC}}{N_{\mu}})\\
(1+\beta_{m}/2)(1-\frac{\rho}{2})\\
\end{array}
\right\}
\end{eqnarray*}

The meaning of the various $\alpha$, $\delta$, $\eta$, $\beta$, $\lambda$ and
$\rho$ are given in table \ref{syst}.
The choice of term from within the large braces depends on which portion of
the data is being compared.

There are a number of problems with this fit.  In particular the expression
$(N^{i}_{data}-N^{i}_{MC})^{2}$ includes a substantial systematic error in
$N^{i}_{MC}$ which is not included in the denominator $\sigma^{2}_{i}$.
Including this error would tend to {\em lower} the $\chi^{2}$.

But the systematic error in $N^{i}_{MC}$ is common to many terms of the sum.
In fact a definition of $\chi^{2}$ including this common error would need to
be of the matrix form:

\[
\overline{{(N_{data}-N_{MC})}}^{T}(\sigma^{2})^{-1} \overline{(N_{data}-N_{MC})}
\]

Where $\overline{{(N_{data}-N_{MC})}}$ is a vector of all of
the differences
and $(\sigma^{2})^{-1}$ is the weight matrix, the inverse of the covariance
matrix, that includes the correlated errors.

Even with these definitions it would be hard to interpret the resulting
$\chi^{2}$ in terms of a confidence level since the {\em a priori}
distributions of the systematic errors are not known.

\subsection{The Asymptotic Assumption}
The fit mentioned in the section above includes an unjustified assumption.
It is assumed that the ``systematic'' parameters may be fit in an unbiased way.
Many of these parameters are correlated with quantities of physical interest.
For example the up to down flux ratio, $\eta_{s}$ is closely tied to any
observed angular asymmetry.  The muon to electron neutrino flux ratio,
$\beta_{s}$ is essentially the denominator if the $R$ value.
The use of $\beta_{s}$ and $\beta_{m}$ effectively decouple the fit from the
measured value of $R$.

The fit assumes that the systematic parameters may be fit to some portion of
the solid angle where the data sample is not anomalous.  In essence the fit
assumes that the upper hemisphere represents a good sample of unoscillated
events.  But as discussed in section II the value of R makes this assumption
rather risky.  Any $R$ interpreted solely in terms of muon neutrino
oscillations must include the upper hemisphere.  On the other hand, $R$ could
be anomalously low because the electron sample is not as expected.


The angular fit is relatively insensitive to a $\Delta m^{2}$ in the range of
$5 \times 10^{-2}$ eV$^{2}$ to $10^{-4}$ eV$^{2}$.  This is because the data
has been collected from primarily
two baselines separated by 3 orders of magnitude.  $\Delta m^{2}$'s over a
rather broad range will give comparable fits to the data.  It is only the
relatively small amount of data near the horizon that is sensitive to these
intermediate ranges of $\Delta m^{2}$.  In general it is easier to study such
ranges by looking at the {\em energy} dependence after, perhaps, integrating
over all solid angle, as we have done in the earlier section of this note.

\section{Time Variation of the Event Rate and $R$}
The event rate and the measured value of $R$ for data collected during most of
1998\cite{mess} was significantly lower than that used in earlier Super
Kamioka papers
on atmospheric neutrinos\cite{vary}.  In particular the overall event rate
during the 321.6 live days of this period dropped by 12$\pm$3\% as compared to
the event rate noted in the earlier 414.4 day period.  A careful comparison
of these newer
data indicates that the muon event rate did not change.  The 1998 muon rate was
3$\pm$5\% above the earlier rate.  But both the electron sample and the
multiprong event rate declined sharply.  The electron sample was down by
18$\pm$5\%.  Table \ref{tab1} summarizes the differences between the 1998
sample the earlier sample.

\begin{table*}
\begin{tabular}{lrrr}
& Cumulative\cite{mess} & Phys. Lett.\cite{sub} & 1998 Sample \\ \hline
Single Ring & 3224 & 1883 & 1341 \\
e-like & 1607 & 983 & 624 \\
$\mu$-like & 1617 & 900 & 717 \\
Multi-Ring & 1271 & 784 & 487 \\
Total & 4495 & 2667 & 1828 \\
$\mu/e$ & 1.01$\pm$0.04 & 0.92$\pm$0.04 & 1.15$\pm$0.06 \\
$R=(\mu/e)_{Obs}/(\mu/e)_{MC}$ & 0.67$\pm$0.02 & 0.61$\pm$0.03
& 0.76$\pm$0.04 \\
Exposure (days) & 736 & 414.4 & 321.6 \\
Event Rate & 6.11$\pm$0.09 & 6.44$\pm$0.12 & 5.68$\pm$0.13 \\
Single Ring Rate & 4.38$\pm$0.08 & 4.54$\pm$0.10 & 4.17$\pm$0.11 \\
e rate & 2.18$\pm$0.05 & 2.37$\pm$0.08 & 1.94$\pm$0.08 \\
$\mu$ rate & 2.20$\pm$0.05 & 2.17$\pm$0.07 & 2.23$\pm$0.08 \\
Multi-Ring Rate & 1.73$\pm$0.05 & 1.89$\pm$0.07 & 1.51$\pm$0.07 \\
\end{tabular}
\caption{\label{tab1}Comparison of Super Kamioka Sub-GeV Samples}
\end{table*}

The change in the measured electron rate means that $R$ must have changed.
$R$ rose from 0.61$\pm$0.03 to 0.76$\pm$0.04.  This rise in $R$ brings it from
a region which is not consistent with the hypothesis of neutrino oscillations
over approximately half of the solid angle to a value consistent with such
a hypothesis.  Unfortunately there is no good reason to doubt the credibility
of either data sample.  Super Kamioka has combined these samples in their
analyses.  The oscillation analysis utilizes the electron events, which are
not consistent through this period.

$R$ depends on the ratio of observations to simulations.  The simulations
did not change in any significant way between these two periods\cite{vary}

So there is clear evidence that the electron sample does not provide a
reliable normalization for the analysis of atmospheric neutrino data.
The assumption that the ratio of atmospheric electron to muon neutrino flux
can be calculated to an accuracy of 5\% is in error, unless some portion of
the observed signal is not of atmospheric origin or is not
neutrinos.\cite{vary}

The original Super Kamioka paper on the atmospheric neutrino flux\cite{sub}
notes,
``Using a detailed Monte Carlo simulation, the ratio 
$\frac{(\mu/e)_{Data}}{(\mu/e)_{MC}}$ was measured to be
0.61$\pm$0.03({\em stat})$\pm$0.05({\em sys}), consistent with previous
results from the Kaimiokande, IMB and Soudan-2 experiments, and smaller
than expected from theoretical models of atmospheric neutrinos''.  This
is no longer true.  The 1998 data is not
consistent with previous results of the Super Kamioka detector or any of the
earlier experiments cited.  An $R$ value of 0.76$\pm$0.04 seems to be
unique to this period.

The multi-GeV data sample has no evidence of a rate variation in 1998.

Independent confirmation of the drop in event rate would help ally fears that
it is due to
instrumental or systematic effects.  The Super Kamioka group has maintained
very good monitoring of the detector performance \cite{kasuga} so such an
explanation seems unlikely.

\section{Conclusions}

Several factors indicate that the atmospheric neutrino anomaly may be
more complicated than generally thought.  In such cases, it is important to
seek corroboration of the results from other experiments.  It is also important
to be aware of simplifying assumptions that may have gone into the
interpretation or the analysis of the data.  One must be careful not to
ignore information simply because it is in conflict with a preferred
solution.

It would appear that the electron neutrinos can no longer be used for
relative normalization, at least until the variation is understood.
In the early days the anomaly was considered to be either a deficiency of muon
like events or an excess of electron type events.  In fact, the results
looked a bit like a systematic error in the classification method used for
electron and muon events.  Since a random error would move events from the
larger population, the muons, to the smaller population, the electrons.
The observed event rate for all events was much closer to most flux models
than might be expected based the errors quoted for the estimates.
Studies using the independent muon decay signature corroborate the
morphology based results for muon and electron classification.

The ``neutrino'' events observed in underground detectors do not agree with
expectations based on a simple source of atmospherically produced hadron and
muon decays.  Uncertainties in these expectations does not provide much
guidance as to what might be wrong.  Simple one component muon neutrino
oscillations are unable to provide a convincing explanation for the small
value of $R$ nor for recent apparent drop in the interaction rate.

\section{Acknowledgements}
I would like to thank David Casper for useful comments on an earlier
version of this manuscript.



\end{document}